\documentclass[twocolumn,showpacs,preprintnumbers,amsmath,amssymb,prl,superscriptaddress]{revtex4-1}

\usepackage{multirow}
\usepackage{amsfonts}
\usepackage{amsopn}
\usepackage[english]{babel}
\usepackage[T1]{fontenc}
\usepackage{times}
\usepackage{mathrsfs}
\usepackage{graphicx}
\usepackage{dcolumn}
\usepackage{bm}
\usepackage[colorlinks,bookmarks=true,citecolor=blue,linkcolor=red,urlcolor=blue]{hyperref}
\usepackage[tight, nooneline]{subfigure}

\begin{document}

\title{Exotic Non-Abelian Topological Defects in Lattice Fractional Quantum Hall States}

\author{Zhao Liu}
\affiliation{Dahlem Center for Complex Quantum Systems and Institut f\"ur Theoretische Physik, Freie Universit\"at Berlin, Arnimallee 14, 14195 Berlin, Germany}
\author{Gunnar M\"oller}
\email{Corresponding author: G.Moller@kent.ac.uk}
\affiliation{Functional Materials Group, School of Physical Sciences, University of Kent, Canterbury CT2 7NZ, United Kingdom}
\affiliation{TCM Group, Cavendish Laboratory, University of Cambridge, Cambridge CB3 0HE, United Kingdom}
\author{Emil J. Bergholtz}
\affiliation{Department of Physics, Stockholm University, AlbaNova University Center, 106 91 Stockholm, Sweden}

\date{\today}

\begin{abstract}
We investigate extrinsic wormholelike twist defects that effectively increase the genus of space in lattice versions of multicomponent fractional quantum Hall systems. Although the original band structure is distorted by these defects, leading to localized midgap states, we find that a new lowest flat band representing a higher genus system can be engineered by tuning local single-particle potentials. Remarkably, once local many-body interactions in this new band are switched on, we identify various Abelian and non-Abelian fractional quantum Hall states, whose ground-state degeneracy increases with the number of defects, i.e, with the genus of space. This sensitivity of topological degeneracy to defects provides a ``proof of concept'' demonstration that genons, predicted by topological field theory as exotic non-Abelian defects tied to a varying topology of space, do exist in realistic microscopic models. Specifically, our results indicate that genons could be created in the laboratory by combining the physics of artificial gauge fields in cold atom systems with already existing holographic beam shaping methods for creating twist defects.
\end{abstract}


\maketitle
\paragraph{Introduction.}Extrinsic defects embedded in topologically ordered phases of matter \cite{laughlin83,senthil06,wen91,moore91,maciejko15} may acquire exotic properties \cite{TCtwists,kane10,freedman11,ady12,you12,qi12,qi13,qi13_2,vaezi13,yy13,brown13,vaezi14,fuchs14,qi15,jiang15,yy16,buehler14}. Genons \cite{qi12,qi13}, named after their ability to effectively increase the genus of space thus enhancing the topological degeneracy, are particularly intriguing representatives of this idea and can be visualized as twist defects at the ends of branch cuts connecting separate ``world sheets'' of different components in the host system. Importantly, the linkage of genons to the topology of space and the underlying topological order establishes them as powerful tools to overcome the long-standing challenge of accessing topological orders on surfaces with tunable genus. It also imparts them with nontrivial quantum dimensions and braiding statistics that are significantly different from those of intrinsic quasiparticles of the host system \cite{qi13}, thus enabling fault tolerant topological quantum computation \cite{kitaev03,topocompute} even in Abelian host states without this capability and extending our knowledge of topological order.  However, while the beautiful idea of genons is based on topological field theory \cite{qi12,qi13} 
and corroborated by complicated exactly solvable models \cite{TCtwists,you12,brown13}, its actual relevance to realistic microscopic models has remained open. 

In this Letter, we fill this void by presenting the first evidence of genons in a microscopic lattice model which can favor lattice fractional quantum Hall states, i.e., fractional Chern insulators \cite{otherreview,Emilreview}, and naturally host defects. With a scheme to offset the negative influence of defects on the band structure, we obtain compelling results that explicitly demonstrate the remarkable fingerprint of genons --- the nontrivial dependence of the topological degeneracy on the number of defects which effectively tune the genus of space to high numbers. Our results provide a deep insight into the physical realization of genons in simple lattice models involving only single-particle hopping and on site two-body interactions, thus opening up the experimental accessibility of topological orders on high-genus surfaces.


\begin{figure}
\centerline{\includegraphics[width=\linewidth]{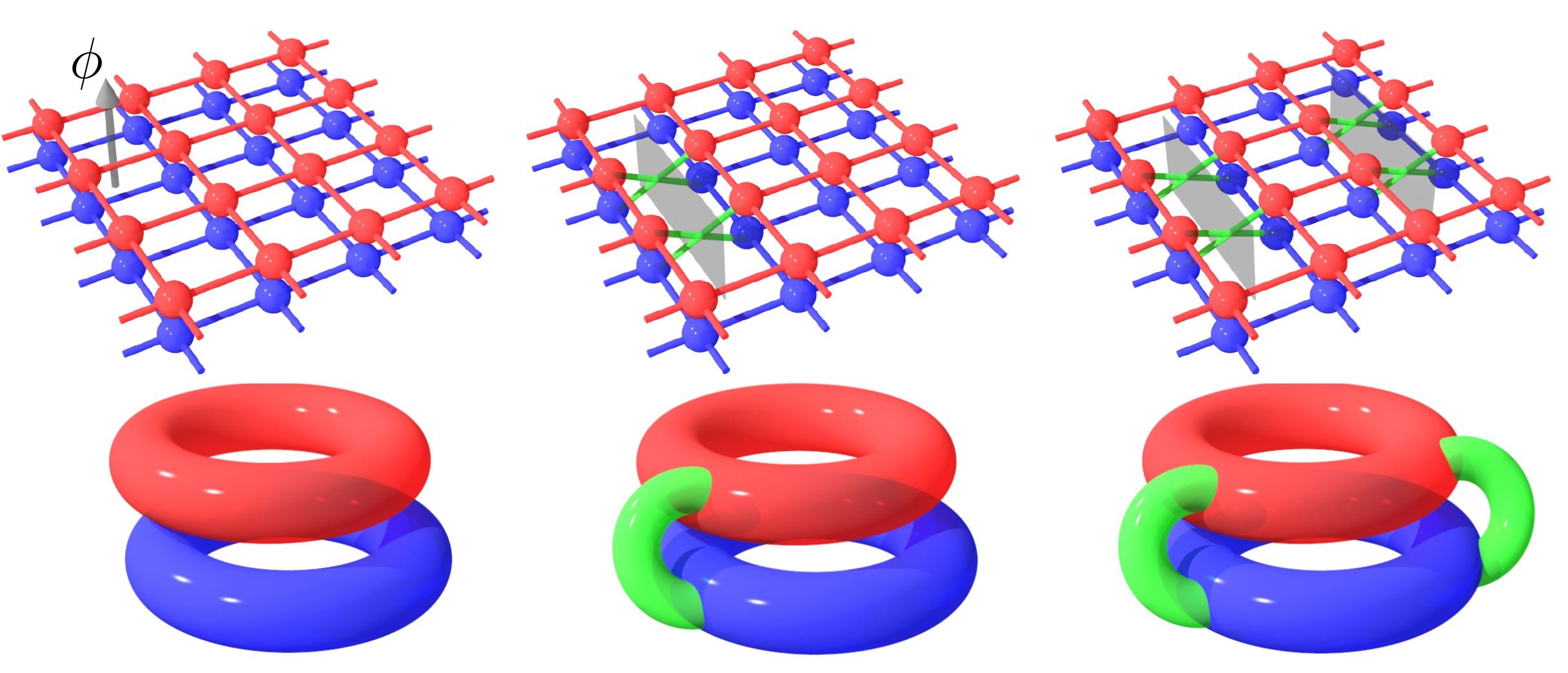}}
\caption{Our model is equivalent to two square lattice layers (blue and red) where each plaquette is pierced by an effective flux $\phi$ (upper left panel). We only plot nearest-neighbor hopping for simplicity.  Defects are introduced through branch cuts (transparent gray) where the particles switch layer (green). We study systems with up to two such branch cuts, corresponding to topologies resembling wormholes, as displayed in the bottom panels.
}
\label{fig:lattice}
\end{figure}


\begin{figure}
\centerline{\includegraphics[width=\linewidth]{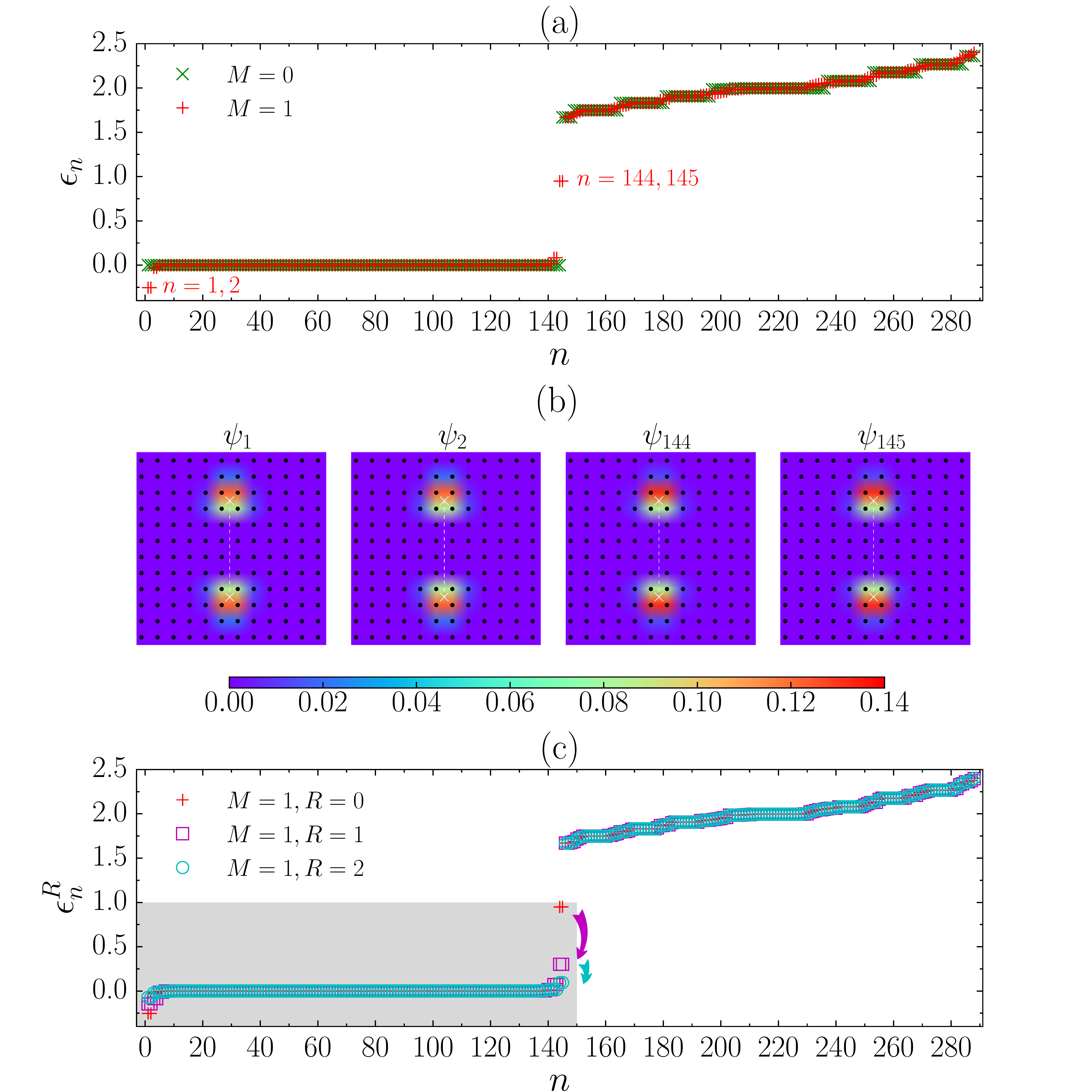}}
\caption{Band structure for a $L_x\times L_y=12\times12$ lattice with $\phi=1/2$. (a) The spectrum $\{\epsilon_n\}$ of $H_0$. Without defects ($M=0$), $\epsilon_{1},\cdots,\epsilon_{144}$ are exactly degenerate at zero energy. With one branch cut [$M=1$, white dashed line in (b)] at $(5.5,2.5\rightarrow 8.5)$, the original band structure is distorted, with two nearly degenerate clusters $(\epsilon_1,\epsilon_2)$ and $(\epsilon_{144},\epsilon_{145})$ having the largest deviation. (b) The lattice site weight of eigenvectors $\psi_1,\psi_2,\psi_{144},\psi_{145}$ of $H_0$ for the same defects as in (a). All of them are strongly localized near the defects. The eigenstates with less energy deviation from the original band structure, for example, $\psi_3,\psi_4,\psi_{142},\psi_{143}$, are less localized (not shown here). (c) The spectrum $\{\epsilon_n^R\}$ of $H_0+V$ with $R=0$, $1$, and $2$ and the same defects as in (a). $\epsilon_{1}^R,\cdots,\epsilon_{145}^R$ which we must flatten (shaded in gray) becomes more degenerate for larger $R$, with the flatness $(\epsilon_{2\phi L_xL_y+M+1}^R-\epsilon_{2\phi L_xL_y+M}^R)/(\epsilon_{2\phi L_xL_y+M}^R-\epsilon_{1}^R)\approx 0.6,3.1,9.4$ for $R=0,1,2$.}
\label{fig:SP}
\end{figure}

\paragraph{Model.}We consider particles in a two-dimensional square lattice 
with two internal degrees of freedom (referred to as ``layers'' for convenience) $\sigma=\uparrow,\downarrow$ on each lattice site and an effective magnetic flux $\phi$ piercing each elementary plaquette (Fig.~\ref{fig:lattice}). We introduce $\mathbb{Z}_2$ twist defects~\cite{qi13} into the lattice such that a particle's layer index is flipped when it moves around such a defect once. It is helpful to imagine that the layer flipping occurs precisely when a particle hops across a branch cut that we take to connect a pair of defects in a straight line (Fig.~\ref{fig:lattice}).
We thus formulate the tight-binding Hamiltonian as
\begin{eqnarray}
\label{H0}
H_0=\sum_{j,k,\sigma}t(z_j,z_k) a^\dagger_{j,\mathcal{F}^{n_{jk}}(\sigma)}a_{k,\sigma},
\end{eqnarray}
where $a^\dagger_{j,\sigma}$ ($a_{j,\sigma}$) creates (annihilates) a particle in
layer $\sigma$ at lattice site $z_j=x_j+\textrm{i}y_j$, and $\mathcal{F}^{n_{jk}}(\sigma)$ accounts for $n_{jk}$ flips of the initial layer $\sigma$ when a straight line from $z_k$ to $z_j$ intersects with $n_{jk}$ branch cuts. The hopping coefficient from $z_k$ to $z_j$ is designed as $t(z_j,z_k)=(-1)^{x+y+xy} e^{-\frac{\pi}{2}(1-\phi)|z|^2} e^{-\textrm{i}\pi\phi(x_j+x_k)y}$~\cite{kapit,tjk}, where $z = z_j - z_k=x+\textrm{i}y$. Such a hopping is local in
the sense that $t(z_j,z_k)$ follows a superexponential decay. We focus on $\phi=1/q$ with integer $q$, for which a unit cell contains $q$ sites in the $x$ direction.
Without defects, $H_0$ has a $\mathbb{Z}_2$ symmetry associated with exchanging two layers and corresponds to two decoupled Kapit-Mueller
models~\cite{kapit} in the Landau gauge; thus, its lowest band contains two copies of an exactly flat band with Chern number $\mathcal{C}=1$.

The effective topology of our model strongly depends on the number of branch cuts (Fig.~\ref{fig:lattice}).
If each layer has a torus geometry,
a branch cut plays the role of a wormhole connecting two tori~\cite{qi12}; hence, $M$ branch cuts effectively lead to a single surface with genus $g=M+1$.
In the following, we arrange all branch cuts in the $y$ direction without loss of generality \cite{supplement}, denoting the branch cut connecting a pair of defects at $(X_1,Y_1)$ and $(X_1,Y_2)$ as $(X_1,Y_1\rightarrow Y_2)$ \cite{branchcut}.

\paragraph{Single-particle spectra and defect-induced localized states.}We diagonalize $H_0$ on a periodic lattice $\mathcal{L}$ of $L_x\times L_y$ sites to analyze the effect of defects on the band structure \cite{Lx}. Without defects, the lowest $2\phi L_xL_y$ single-particle levels are exactly degenerate at zero energy. This flatness is seriously distorted by $M$ pairs of defects, and we identify $4M$ levels with a significant deviation from the original band structure: $2M$ of them (levels $\epsilon_1,...,\epsilon_{2M}$) drop below the original lowest band, and another $2M$ ($\epsilon_{2\phi L_xL_y-M+1},...\epsilon_{2\phi L_xL_y+M}$) move into the original lowest band gap. Moreover, they form nearly degenerate clusters, respectively. An example of the band structure for $M=1$ is shown in Fig.~\ref{fig:SP}(a). We further examine the eigenvectors of these $4M$ levels. Remarkably, they are all strongly localized near the defects [Fig.~\ref{fig:SP}(b)], and the localization becomes weaker or completely disappears for other levels with less deviation from the original band structure. This localization enables us to do a controlled tuning of the deviated energies by local potentials near the defects without significantly distorting the rest of the band structure, as we explain below.

\paragraph{Higher genus flat bands.}The effectively increased genus does not guarantee that defects in our model can be thought of as genons. We must show that topological phases can be stabilized on the high-genus surfaces created by these defects, and that they display defect-enhanced topological degeneracy. Tuning deviated single-particle energies to recover a flat lowest band is necessary for reaching this goal. We consider $N_b=\frac{k}{2}(2\phi L_xL_y)$ bosons interacting via $(k+1)$-body on site repulsions
\begin{eqnarray}
\label{Hint}
H_{\mathrm{int}}=\sum_{i\in\mathcal{L},\sigma=\uparrow,\downarrow} :n_{i,\sigma}n_{i,\sigma}\cdots n_{i,\sigma}:
\end{eqnarray}
with integer $k\geq 1$ \cite{hafezi14}. In this setup, the ground state without defects is two copies of model $\mathcal Z_k$ Read-Rezayi (RR) states on the lattice, residing in the lowest $2\phi L_xL_y$ exactly degenerate eigenstates of $H_0$ with filling fraction $\nu=N_b/(2\phi L_xL_y)=k/2$ \cite{zhaononabelian,ModelFCI}. Adding $M$ pairs of defects effectively deforms the topology to a single $g=M+1$ surface but should not change $\nu$ in the thermodynamic limit. Hence, in that case the most promising candidate for the underlying topological phase is the $\mathcal Z_k$ RR state on a single $g=M+1$ surface. In the continuum, such a state resides in $N_s$ exactly degenerate single-particle states in the lowest Landau level, with
\begin{eqnarray}
\label{Sg}
N_s=2N_b/k-(1-g),
\end{eqnarray}
where $\nu=\lim_{N_b\rightarrow\infty}N_b/N_s=k/2$, and the extra offset $1-g$ is related to the topological ``shift'' \cite{wenshift,eq3}. Consequently, in our lattice model with $M$ pairs of defects, Eq.~(\ref{Sg}) combined with $N_b=\frac{k}{2}(2\phi L_xL_y)$ and $g=M+1$ requires a flat band consisting of the lowest $N_s=2\phi L_xL_y+M$ single-particle eigenstates of $H_0$ to host the $\mathcal Z_k$ RR state. However, this set of eigenstates corresponds to a residual flat band plus all significantly deviated levels [Fig.~\ref{fig:SP}(a)]. As the emergence of FQH liquids requires a hierarchy of energy scales such that interactions dominate the band dispersion of the low-energy band, we must first flatten this large band dispersion to amplify the interaction effect before a topological state can be realized. Fortunately, this can be readily achieved by local potentials owing to the strong localization of the deviated states near defects [Fig.~\ref{fig:SP}(b)]. A simple candidate of such a local potential \cite{supplement} is $V=-\sum_{n=1}^{2\phi L_xL_y+M} \epsilon_n \mathcal{T}_R(|\psi_n\rangle\langle \psi_n|)$, where $\epsilon_n$'s and $\psi_n$'s are the eigenvalues and eigenvectors of $H_0$, respectively, and $\mathcal{T}_R$ denotes the truncation at a radius $R$ around each defect. The dominant terms in $V$ exactly correspond to the deviated levels, because others staying at $\epsilon_n=0$ have no contributions.
As expected, a very small $R$ is already sufficient to do the flattening very well, with negligible influence on the pertinent eigenvector subspace. In Fig.~\ref{fig:SP}(c), we show the band structure of $H_0+V$ with $M=1$ and $R=0,1,2$, respectively. The degeneracy between the lowest $2\phi L_xL_y+M$ energy levels indeed becomes better with the increase of $R$, with the flatness significantly increased to $\approx 9.4$ for $R=2$. The corresponding eigenvectors of $H_0+V$ have a total $99\%$ overlap with those of $H_0$ for $R=1$ and $R=2$.

\begin{figure}
\centerline{\includegraphics[width=\linewidth]{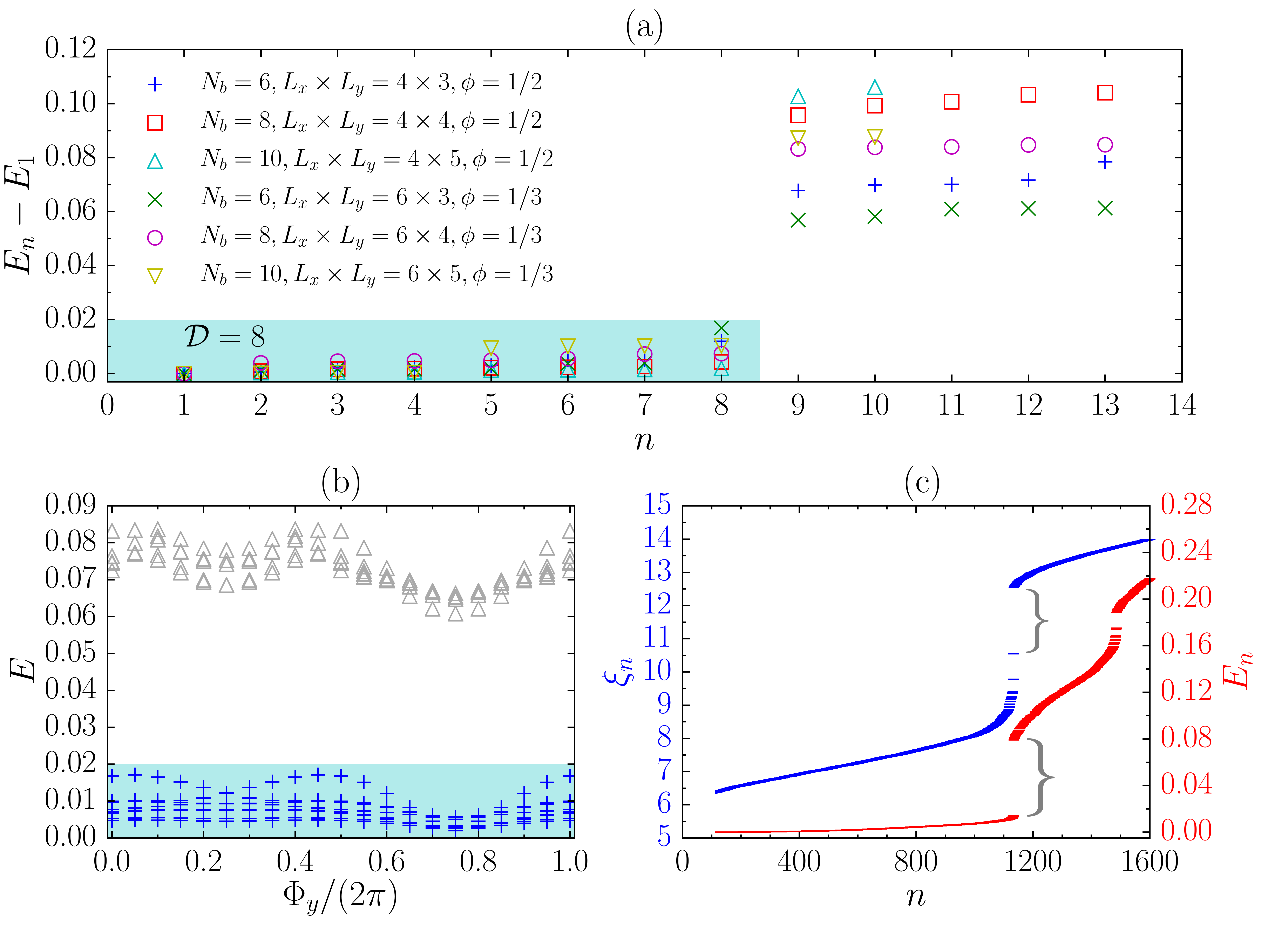}}
\caption{Defect-enhanced topological degeneracy for Abelian systems at $\nu=1/2$ with two branch cuts \cite{branchcut}. (a) The energy spectra of various system sizes. The eight quasidegenerate ground states are highlighted by the cyan shade. (b) The $y$-direction spectral flow for $N_b=6,L_x\times L_y=4\times 3,\phi=1/2$. The eight ground states (blue $+$) never mix with excited states (gray $\triangle$). (c) The PES (blue) for $N_b=8,L_x\times L_y=6\times 4,\phi=1/3$ in the $N_b^A=4$ sector and the corresponding quasihole excitations (red) for $N_b=4,L_x\times L_y=6\times 4,\phi=1/3$. The number of states below the gaps (indicated by the gray $\}$) are identical in both spectra.}
\label{fig:Lau}
\end{figure}

\paragraph{Defect-enhanced topological degeneracy.}After ensuring that a new lowest flat band can be recovered, we are now in the position to examine whether interactions can stabilize the $\mathcal Z_k$ RR states in the single high-genus surfaces created by defects, characterized by the defect-enhanced topological degeneracy $\mathcal{D}$ \cite{Zij}. We project the interaction $H_{\mathrm{int}}$, which is assumed to be small relative to the band gap, to the lowest $2\phi L_xL_y+M$ eigenstates of $H_0$ \cite{projection} and neglect their energy dispersion for large numerical efficiency. This procedure is similar to the band projection in the flat-band limit extensively used to study fractional Chern insulators without defects \cite{rbprx}.

In the most realistic $k=1$ case, we find compelling evidence that defects lead to a $\nu=1/2$ Laughlin state on effective high-genus surfaces. Without defects, the ground state is two copies of $\nu=1/2$ Laughlin states on the torus with $\mathcal{D}=2\times 2=4$. Although we still get $\mathcal{D}=4$ with one pair of defects, consistent with the $\nu=1/2$ Laughlin state on a single $g=2$ surface, a nontrivial enhancement of $\mathcal{D}$ from $4$ to $8$ occurs for two pairs of defects ($g=3$), characterized by eight approximately degenerate ground states
for various system sizes [Fig.~\ref{fig:Lau}(a)]. These states are separated from other excited states by an energy gap which is significantly larger than the ground-state splitting, and the splitting is reduced relative to the gap as the system size -- and thus the separation of defects -- is increased. The eight ground states never mix with other excited states under twisted boundary conditions \cite{tjk} [Fig.~\ref{fig:Lau}(b)], which confirms the robustness of topological degeneracy. In order to further corroborate their topological nature, we compute the particle entanglement spectra (PES) \cite{pes, rbprx, peslattice} to probe the quasihole excitation property. We find a clear gap in the PES, at the number of levels matching the corresponding counting of quasihole excitations [Fig.~\ref{fig:Lau}(c)] \cite{rbprx,supplement}. Our results unambiguously indicate that the ground state with $M$ pairs of defects is the $\nu=1/2$ Laughlin state on a single $g=M+1$ surface with degeneracy $\mathcal{D}_M^{k=1}=2^{M+1}$. 
While the inclusion of a local potential $V$ is crucial for obtaining topological degeneracies, the specific choice thereof is less crucial for larger systems stemming from their topological origin \cite{supplement}.

\begin{figure}
\centerline{\includegraphics[width=\linewidth]{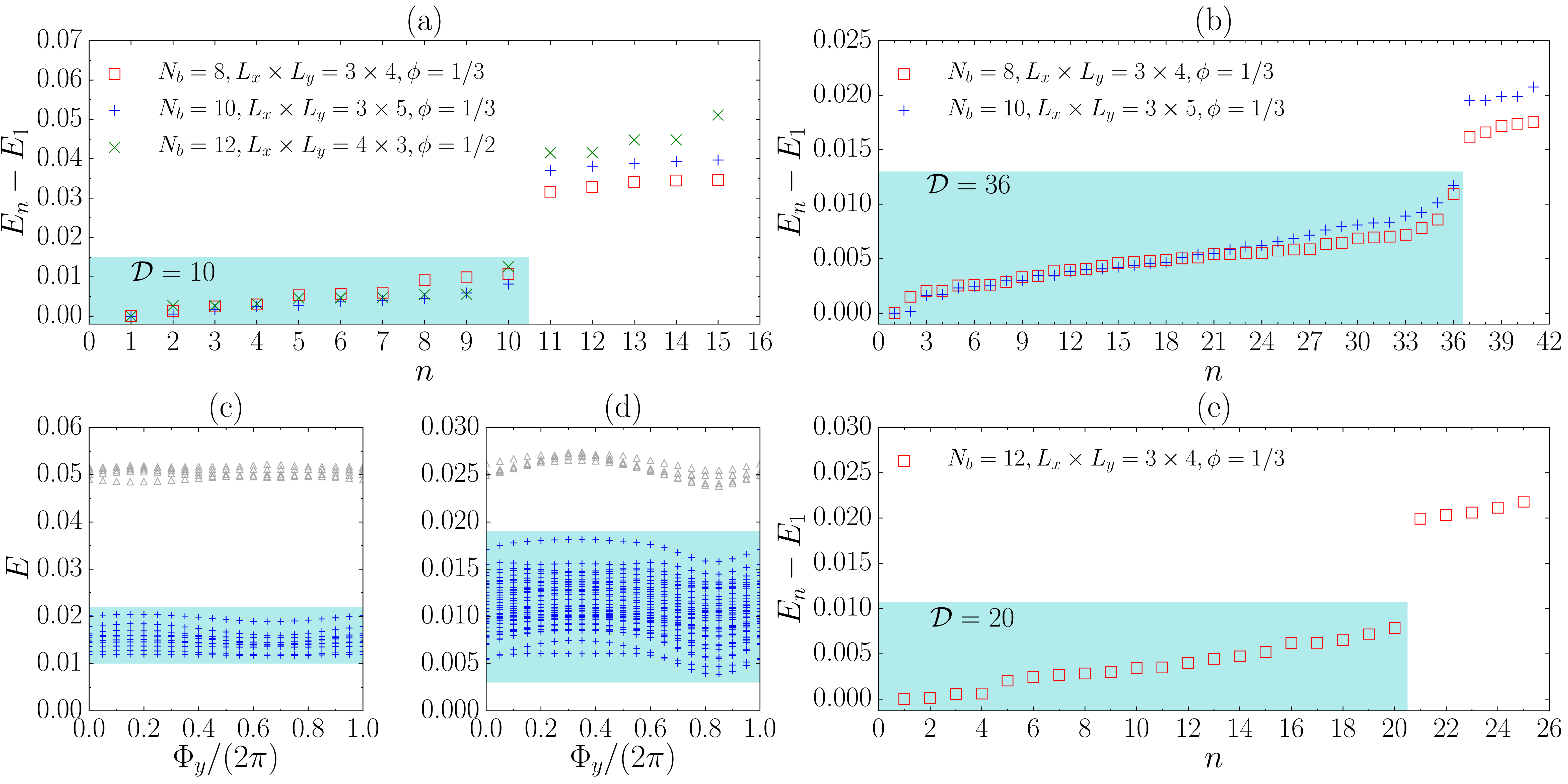}}
\caption{Defect-enhanced topological degeneracy for non-Abelian systems. The approximately degenerate ground states, together with the degeneracy $\mathcal{D}$, are highlighted by the cyan shade. (a) The energy spectra at $\nu=1$ with one pair of defects \cite{branchcut}. (b) The energy spectra at $\nu=1$ with two pairs of defects \cite{branchcut}. (c) The $y$-direction spectral flow for $N_b=10,L_x\times L_y=3\times 5,\phi=1/3$ with the same branch cut as in (a). (d) The $y$-direction spectral flow for $N_b=10,L_x\times L_y=3\times 5,\phi=1/3$ with the same branch cuts as in (b). (e) The energy spectrum at $\nu=3/2$ with one pair of defects \cite{branchcut}.}
\label{fig:nA}
\end{figure}

The effect of defects is even more intriguing at higher $k$'s with non-Abelian host states. For $k=2$, the ground state in the absence of defects is two copies of Moore-Read (MR) states on the torus, with $\mathcal{D}=9$ for even $N_b/2$ and $\mathcal{D}=1$ for odd $N_b/2$. Strikingly, in this case, unlike the situation of $k=1$, one pair of defects already leads to a nontrivial enhancement of $\mathcal{D}$ to $10$ for all even $N_b$, which becomes better for larger system sizes and is robust under twisted boundary conditions [Figs.~\ref{fig:nA}(a) and \ref{fig:nA}(c)]. By adding another pair of defects, $\mathcal{D}$ is further enhanced to $36$ [Figs.~\ref{fig:nA}(b) and \ref{fig:nA}(d)], with a faster growth rate than the $k=1$ case. The dependence of the topological degeneracy on the number of defects convincingly suggests that, by introducing $M$ pairs of defects for $k=2$, the ground state evolves to the $\nu=1$ MR state on a single $g=M+1$ surface with degeneracy $\mathcal{D}_M^{k=2}=2^{M}(2^{M+1}+1)$ \cite{Zkdeg}. The enhancement of the topological degeneracy is also observed for $k=3$, where $\mathcal{D}$ is increased from $16$ to $20$ by adding $M=1$ pair of defects [Fig.~\ref{fig:nA}(e)], consistent with the $\nu=3/2$ $\mathcal{Z}_3$ RR state on a single $g=M+1$ surface with degeneracy $\mathcal{D}_M^{k=3}=2[(5+\sqrt{5})^{M}+(5-\sqrt{5})^{M}]$ \cite{Zkdeg}.

The topological phases with defect-enhanced ground-state degeneracy strongly indicate that the defects in our model are indeed genons.
In particular, each of them carries a distinct nontrivial quantum dimension $d=\lim_{M\rightarrow\infty}(\mathcal{D}_M)^{\frac{1}{2M}}$
from that of intrinsic quasiparticles of the host state.
At $\nu=1/2$, we have non-Abelian genons with $d=\sqrt{2}$, although the Laughlin state only has Abelian quasiparticles. More saliently, genons at $\nu=1$ in our model have $d=2$ thus allowing for universal quantum computation, while the quasiparticles of the MR state itself cannot \cite{topocompute,qi13}. At $\nu=3/2$, we obtain genons with even higher quantum dimension $d=(5+\sqrt{5})^{1/2}$. These differences, together with the projective braiding statistics of defects \cite{qi13}, open the possibility that genons are more powerful tools for topological quantum computation than ordinary quasiparticles.

\paragraph{Discussion.}In this work, we condense the beautiful idea of genons from topological field theory into a recipe for realistic
microscopic lattice models. We identify a number of different lattice genons in both Abelian and non-Abelian host states based on their numerically observed defect-enhanced ground-state degeneracy, which can be thought of as adding genons into the system.
The key ingredients of our proposal are already experimentally available and their combined synthesis is plausibly within reach --- especially for coupled Laughlin states emerging from a particularly simple on site two-body interaction.
Artificial gauge fields generated by lattice shaking techniques are compatible with multiple internal degrees of freedom as we require. The long-range hopping, which is chosen for theoretical elegance and numerical efficiency, is in fact not essential for the existence of lattice genons \cite{supplement}.
Hence, the already realized Hofstadter model in optical lattices \cite{hofstadterexp1,hofstadterexp2} can serve as an eminently promising candidate platform for creating genons, while its higher Chern bands provide an additional variety of host quantum Hall liquids \cite{HofstadterHigherC09, HofstadterHigherC15}.
In particular, a recent realization \cite{hofstadterexp3} based on a quantum gas microscope already allows single-site addressing, and could be combined with holographic beam shaping methods \cite{microscope} that provide a natural route towards producing the branch cuts and local potentials necessary to realize lattice genons as we envision. Furthermore, a time-dependent control over the locations of such branch cuts would enable braiding experiments that may directly probe their exchange statistics.

\begin{acknowledgments}
\paragraph{Acknowledgments.} We thank N.R.~Cooper and J.~Behrmann for useful discussions. Z.L. was supported by an Alexander von Humboldt Research Fellowship for Postdoctoral Researchers and the U.S. Department of Energy, Office of Basic Energy Sciences through Grant No. DE-SC0002140. The latter was specifically for the use of computational facilities at Princeton
University. G.M. is supported by The Royal Society, Grant No. UF120157, and acknowledges use of the Darwin Supercomputer of the University of Cambridge High Performance Computing Service, funded by Strategic Research Infrastructure Funding from the Higher Education Funding Council for England and funding from the Science and Technology Facilities Council.
E.J.B. was supported by the Swedish research council (VR) and the Wallenberg Academy Fellows program of the Knut and Alice Wallenberg Foundation.

All three authors contributed equally to this work.
\end{acknowledgments}

\section{Supplementary Material}
\setcounter{subsection}{0}
\setcounter{equation}{0}
\setcounter{figure}{0}
\renewcommand{\theequation}{S\arabic{equation}}
\renewcommand{\thefigure}{S\arabic{figure}}
\renewcommand{\thesubsection}{S\arabic{subsection}}
\renewcommand{\bibnumfmt}[1]{[S#1]}
\renewcommand*{\citenumfont}[1]{S#1}

\noindent{\bf Short-range tight-binding Hamiltonian.} In the main text, we use a tight-binding Hamiltonian with local but long-range hopping for theoretical elegance. Now we show that we can obtain similar results with only short-range hopping.

First, we only keep the nearest-neighbor (NN) and next-nearest-neighbor (NNN) hopping in the tight-binding model, Eq.~(\ref{H0}) in the main text. Thus, we obtain a new tight-binding model with short-range hopping
\begin{eqnarray}
H_0'=\sum_{j,k,\sigma}t'(z_j,z_k) a^\dagger_{j,\mathcal{F}^{n_{jk}}(\sigma)}a_{k,\sigma},
\end{eqnarray}
where $t'(z_j,z_k)=(-1)^{x+y+xy} e^{-\frac{\pi}{2}(1-\phi)|z|^2} e^{-\textrm{i}\pi\phi(x_j+x_k)y}$ for $|z|^2\leq 2$
and $t'(z_j,z_k)=0$ for $|z|^2>2$. The meanings of the symbols are the same as those in the main text.

\begin{figure*}[b]
\centerline{\includegraphics[width=\linewidth]{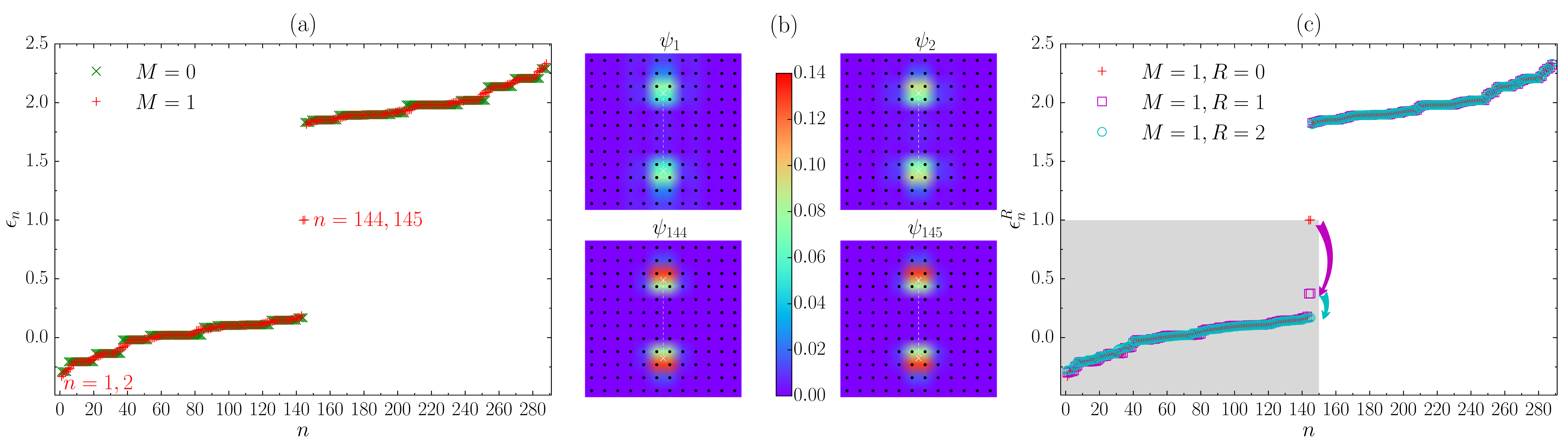}}
\caption{{\bf Single-particle spectra and defect-induced localized states with NN and NNN hopping only.} We study the band structure on a $L_x\times L_y=12\times12$ lattice with $\phi=1/2$. (a) The single-particle spectrum $\{\epsilon_n\}$ of $H_0'$. In the absence of defects ($M=0$), $\epsilon_{1},\cdots,\epsilon_{144}$ are no longer exactly degenerate at zero energy. With a branch cut ($M=1$, white dashed line) at $(5.5,2.5\rightarrow 8.5)$, the original band structure is distorted, with one nearly degenerate cluster $(\epsilon_{144},\epsilon_{145})$ having the largest deviation. (b) The lattice site weight of eigenvectors $\psi_1,\psi_2,\psi_{144},\psi_{145}$ of $H_0'$ for the same defects as in (a). All of them are strongly localized near the defects. However, the localization of $\psi_1$ and $\psi_2$ is weaker than the case of $H_0$ in the main text, probably because now they have much less energy deviation from the original band structure. (c) The single-particle spectrum $\{\epsilon_n^R\}$ of $H_0'+V$ with $R=0$, $1$, and $2$ and the same defects as in (a). The degeneracy of $\epsilon_{1}^R,\cdots,\epsilon_{145}^R$ (shaded in gray) becomes better for larger $R$, with the flatness $0.6,2.2,3.7$ for $R=0,1,2$.}
\label{fig:SP_NNN_1_2}
\end{figure*}

\begin{figure*}
\centerline{\includegraphics[width=\linewidth]{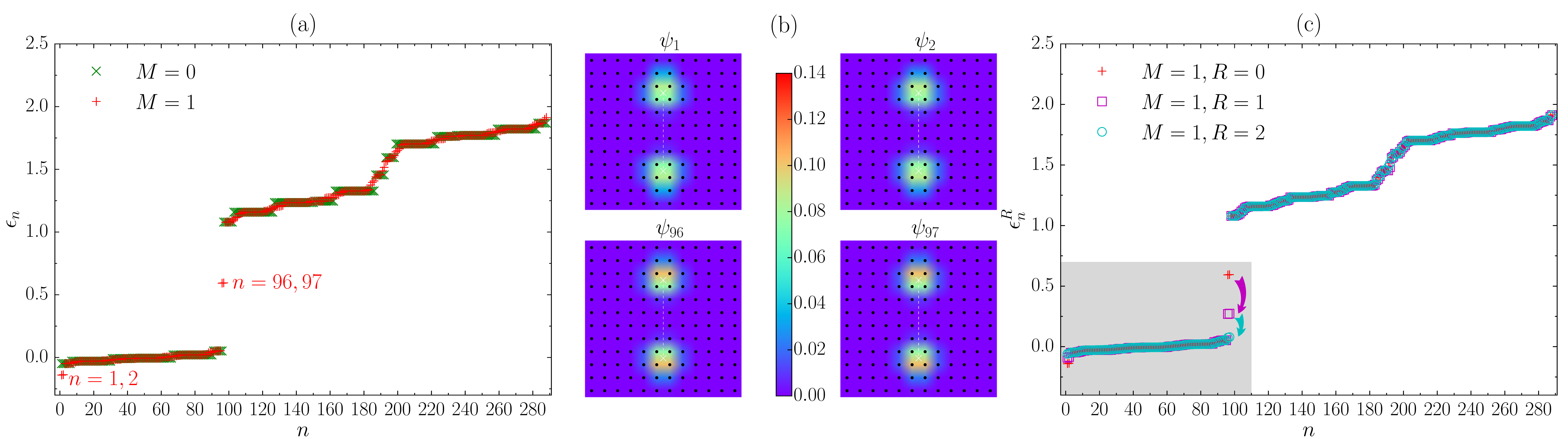}}
\caption{{\bf Single-particle spectra and defect-induced localized states with NN and NNN hopping only.} We show the band structure on a $L_x\times L_y=12\times12$ lattice with $\phi=1/3$. (a) The single-particle spectrum $\{\epsilon_n\}$ of $H_0'$. In the absence of defects ($M=0$), $\epsilon_{1},\cdots,\epsilon_{96}$ are no longer exactly degenerate at zero energy. With a branch cut ($M=1$, white dashed line) at $(5.5,2.5\rightarrow 8.5)$, the original band structure is distorted, with two nearly degenerate clusters $(\epsilon_{1},\epsilon_{2})$ and $(\epsilon_{96},\epsilon_{97})$ having the largest deviation. (b) The lattice site weight of eigenvectors $\psi_1,\psi_2,\psi_{96},\psi_{97}$ of $H_0'$ for the same defects as in (a). All of them are strongly localized near the defects. (c) The single-particle spectrum $\{\epsilon_n^R\}$ of $H_0'+V$ with $R=0$, $1$, and $2$ and the same defects as in (a). The degeneracy of $\epsilon_{1}^R,\cdots,\epsilon_{97}^R$ (shaded in gray) becomes better for larger $R$, with the flatness $0.7,2.2,7.3$ for $R=0,1,2$.}
\label{fig:SP_NNN_1_3}
\end{figure*}

Although the exact flatness of the lowest $2\phi L_xL_y$ eigenstates in the absence of defects is lost due to the hopping truncation, we still find that defects have almost the same effect on the band structure of $H_0'$ as that on $H_0$ shown in the main text [Figs.~\ref{fig:SP_NNN_1_2} and \ref{fig:SP_NNN_1_3}]. The energies of some eigenstates localized near the defects deviate from the original bands, and the dispersion of the lowest $2\phi L_xL_y+M$ eigenstates can be reduced by a local potential $V=-\sum_{n=1}^{2\phi L_xL_y+M} \epsilon_n \mathcal{T}_R(|\psi_n\rangle\langle \psi_n|)$ with negligible influence on the pertinent
eigenvector subspace of $H_0'$, where $\epsilon_n$'s and $\psi_n$'s are now the eigenvalues and eigenvectors of $H_0'$ respectively. One can notice that such a flattening procedure works better for smaller $\phi$ [Figs.~\ref{fig:SP_NNN_1_2} and \ref{fig:SP_NNN_1_3}].

\begin{figure*}
\centerline{\includegraphics[width=0.85\linewidth]{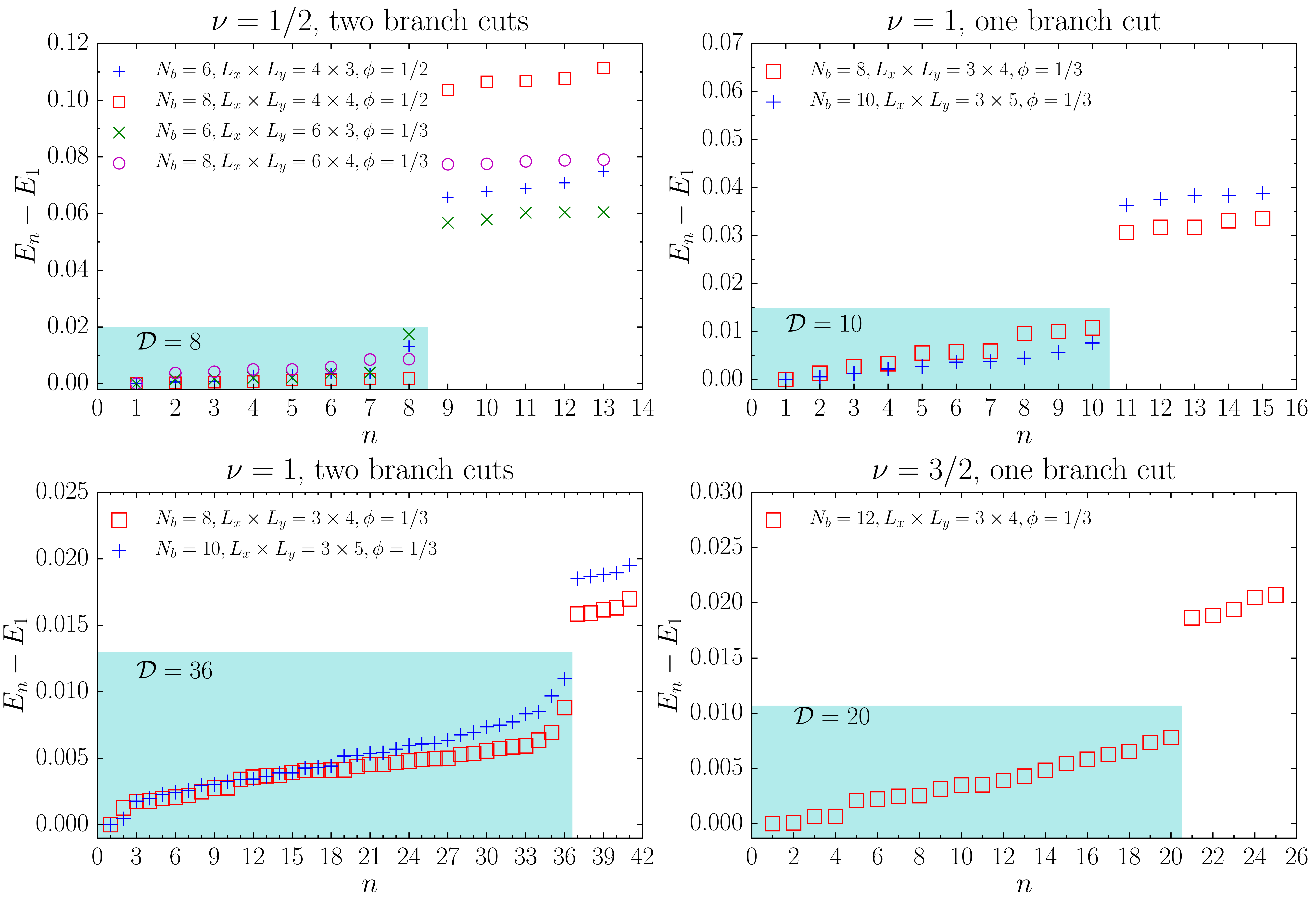}}
\caption{{\bf Defect-enhanced topological degeneracy with NN and NNN hopping only.} We show the many-body spectra resulting from the single-particle Hamiltonian $H_0'$. The interactions and branch cut locations in each specific system size are the same as those used in the main text with $H_0$. The approximately degenerate ground states, together with the degeneracy $\mathcal{D}$, are highlighted by the cyan shade. One can notice that we get the same topological degeneracy as that in the main text.}
\label{fig:TD_NNN}
\end{figure*}

We diagonalize the interaction projected onto the lowest $2\phi L_xL_y+M$ eigenstates of $H_0'$ to examine the topological degeneracy at various filling fractions.
Strikingly, we can get the expected topological degeneracy even though we have truncated the hopping [Fig.~\ref{fig:TD_NNN}].

\begin{figure}
\centerline{\includegraphics[width=\linewidth]{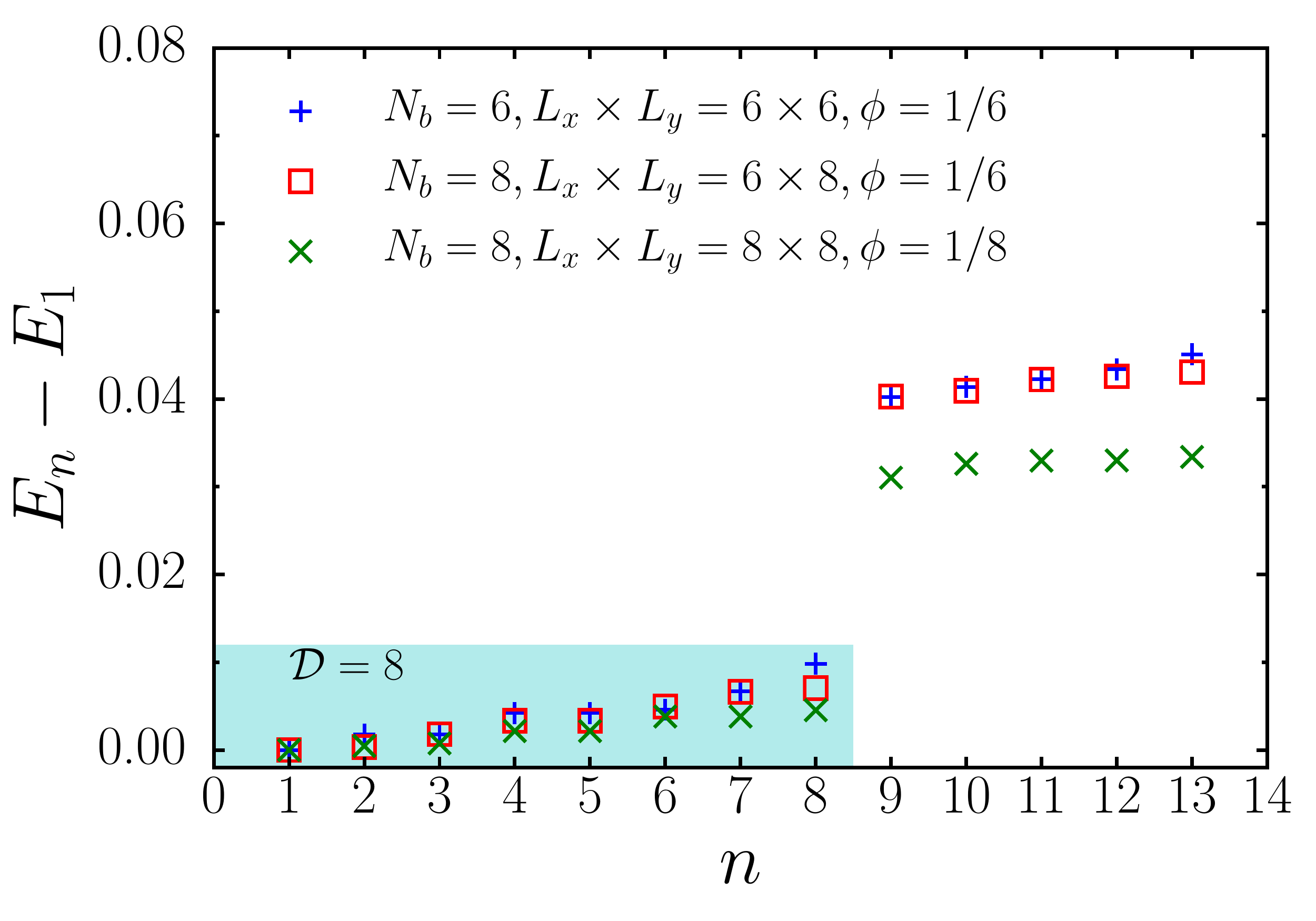}}
\caption{{\bf Defect-enhanced topological degeneracy in the Hofstadter model with defects, for the Abelian $\nu=1/2$ state.}  We show the many-body calculations with NN hopping only for two branch cuts at $\nu=1/2$. Two branch cuts are located at $(0.25,0.5\rightarrow 3.5),(3.25,0.5\rightarrow 3.5)$ for $L_x\times L_y=6\times 6$; $(0.25,1.5\rightarrow 5.5),(3.25,1.5\rightarrow 5.5)$ for $L_x\times L_y=6\times 8$; and $(1.5,1.5\rightarrow 5.5),(5.5,1.5\rightarrow 5.5)$ for $L_x\times L_y=8\times 8$.}
\label{fig:TD_NN}
\end{figure}

Second, let us further truncate the hopping range to include {\em only the nearest-neighbor} terms of the conventional Harper-Hofstadter model \cite{Harper:1955bj,Azbel:1964tk,Hofstadter:1976wt}, with the same type of defects added.
Remarkably, the defect-enhanced eight-fold Laughlin degeneracy of projected interactions remains stable for small flux density $\phi$ even in this case [Fig.~\ref{fig:TD_NN}]. These results imply that the long-range hopping is indeed not necessary for the realization of lattice genons, thus facilitating their experimental realization. A realization based on the nearest-neighbor Harper-Hofstadter model would provide an additional range of host states to explore, as single layers can be chosen to realize higher Chern number $\mathcal{C}$ bands that support a series of hierarchy states at filling factors $\nu = r/(k\mathcal{C}r + 1)$, with $r\in\mathbb{Z}$ and $k$ even (odd) for bosons (fermions) \cite{HofstadterHigherC}.

\noindent{\bf Simplified local potentials.} In the main text, we use an additional potential $V=-\sum_{n=1}^{2\phi L_xL_y+M} \epsilon_n \mathcal{T}_R(|\psi_n\rangle\langle \psi_n|)$  that is localized near the ends of branch cuts in order to restore a flat lowest band. At $R\rightarrow \infty$, this flattening process by $V$ is asymptotically exact in the sense that the lowest $2\phi L_xL_y+M$ eigenstates of $H_0+V$ will become exactly degenerate again at zero energy and have the same eigenvectors as those of $H_0$. Although having an elegant mathematical form, the hopping range in $V$ depends on $R$. In order to facilitate realistic experimental implementations, we now consider a simplified version of $V$ that only contains single-site energies and NN hopping terms: $\widetilde{V}=\alpha\sum_{n=1}^{2M}\mathcal{T}_{R,{\rm NN}}(|\psi_n\rangle\langle \psi_n|)+\beta\sum_{n=2\phi L_xL_y-M+1}^{2\phi L_xL_y+M}\mathcal{T}_{R,{\rm NN}}(|\psi_n\rangle\langle \psi_n|)$.
Here we only sum over the $4M$ single-particle states with the largest deviations from the original band structure (see Sec.~III in the main text). $|\psi_n\rangle$'s are still the eigenvectors of the tight-binding Hamiltonian before band corrections. $\mathcal{T}_{R,{\rm NN}}$ truncates $|\psi_n\rangle\langle \psi_n|$ not only at the radius $R$ around each defect, but also up to the NN hopping. $\alpha$ and $\beta$ are parameters which we need to optimize to pursue the flattest lowest band.

\begin{figure*}
\centerline{\includegraphics[width=\linewidth]{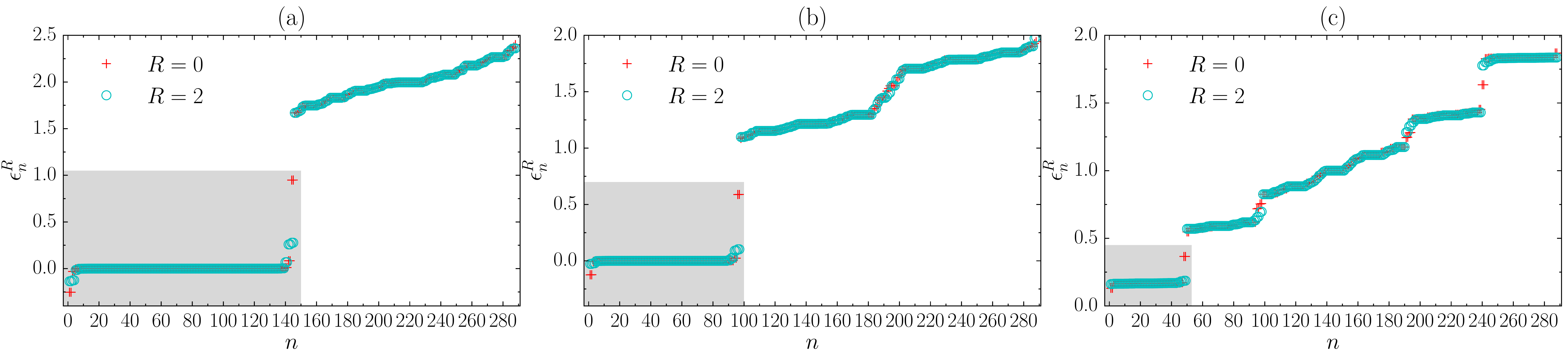}}
\caption{{\bf Flattening the lowest band by a local potential $\widetilde{V}$ including only single-site energies and NN hoppings.} We show the band structure on an $L_x\times L_y=12\times12$ lattice with a single branch cut ($M=1$) at $(5.5,2.5\rightarrow 8.5)$. (a) The single-particle spectrum $\{\epsilon_n^R\}$ of $H_0+\widetilde{V}$ at $\phi=1/2$ with $(R,\alpha,\beta)=(0,0,0)$ and $(2,0.8,-1.4)$. (b) The single-particle spectrum $\{\epsilon_n^R\}$ of $H_0+\widetilde{V}$ at $\phi=1/3$ with $(R,\alpha,\beta)=(0,0,0)$ and $(2,0.9,-1.3)$. (c) The single-particle spectrum $\{\epsilon_n^R\}$ of $H^{{\rm Hof}}+\widetilde{V}$ at $\phi=1/6$ with $(R,\alpha,\beta)=(0,0,0)$ and $(2,1,-1.3)$, where $H^{{\rm Hof}}$ is the conventional Harper-Hofstadter model with added defects. One can see that, compared to the spectrum without $\widetilde{V}$ correction [$(R,\alpha,\beta)=(0,0,0)$], a flat lowest band (shaded in gray) required to stabilize the RR state is indeed established by $\widetilde{V}$, with a flatness ratio of $3.4,7.7$ and $15.5$ in (a), (b) and (c) respectively.}
\label{fig:SP_V}
\end{figure*}

We find that $\widetilde{V}$ with small $R$ is sufficient to flatten the lowest band, with negligible influence on the pertinent
eigenvector subspace of the tight-binding Hamiltonian before band corrections. As shown in Fig.~\ref{fig:SP_V}, a flat lowest band required by the RR state is restored by $\widetilde{V}$ for $H_0$ [Eq.~(1) in the main text] [Figs.~\ref{fig:SP_V}(a) and (b)] as well as for the conventional Hofstadter model with defects at small flux density [Fig.~\ref{fig:SP_V}(c)]. Therefore, $\widetilde{V}$, which only contains single-site and NN terms near each defect, is potentially suitable for the experimental realization of genons. In practice, one can even simply it further by eliminating some terms with small coefficients from $\widetilde{V}$.

\noindent{\bf Straight branch cuts used in the main text.} For completeness, we indicate the precise positions of branch cuts used to generate the data in the main text. All cuts in the main text are oriented along the $y$-axis, and for such branch cuts connecting a pair of two defects with identical $X_1$-coordinate and positions $(X_1,Y_1)$ and $(X_1,Y_2)$ we use the more succinct notation $(X_1,Y_1\rightarrow Y_2)$.

For $\nu=1/2$ with two pairs of defects (Fig.~\ref{fig:Lau}), two branch cuts are located at $(0.5,0.25\rightarrow 1.75),(2.5,0.25\rightarrow 1.75)$ for $L_x\times L_y=4\times 3$; $(0.5,0.5\rightarrow 2.5),(2.5,0.5\rightarrow 2.5)$ for $L_x\times L_y=4\times 4$ and $4\times 5$; $(0.25,0.25\rightarrow 1.75),(3.25,0.25\rightarrow 1.75)$ for $L_x\times L_y=6\times 3$; $(0.25,0.5\rightarrow 2.5),(3.25,0.5\rightarrow 2.5)$ for $L_x\times L_y=6\times 4$; and $(0.25,0.75\rightarrow 3.25),(3.25,0.75\rightarrow 3.25)$ for $L_x\times L_y=6\times 5$. For $\nu=1$ with one pair of defects [Fig.~\ref{fig:nA}(a)], the branch cut is located at $(0.5,0.5\rightarrow 2.5)$ for $L_x\times L_y=3\times 4$; $(0.5,0.75\rightarrow 3.25)$ for $L_x\times L_y=3\times 5$; and $(1.5,0.25\rightarrow 1.75)$ for $L_x\times L_y=4\times 3$. For $\nu=1$ with two pairs of defects [Fig.~\ref{fig:nA}(b)], the branch cuts are located at $(0.25,0.5\rightarrow 2.5),(1.75,0.5\rightarrow 2.5)$ for $L_x\times L_y=3\times 4$ and $(0.25,0.75\rightarrow 3.25),(1.75,0.75\rightarrow 3.25)$ for $L_x\times L_y=3\times 5$. For $\nu=3/2$ with one pair of defects [Fig.~\ref{fig:nA}(e)], the branch cut is located at $(0.25,0.5\rightarrow 2.5)$ for $L_x\times L_y=3\times 4$.

\noindent{\bf Tilted branch cuts.}
In the main text, we have presented data for branch cuts arranged in the $y$-direction, as detailed above. In addition, we now consider more general locations of defects that yield tilted branch cuts. In these general cases,
we denote the branch cut connecting a pair of defects at $(X_1,Y_1)$ and $(X_2,Y_2)$ as $(X_1,Y_1)\rightarrow(X_2,Y_2)$.
In the following we use the same tight-binding model $H_0$ as in the main text.

\begin{figure*}
\centerline{\includegraphics[width=\linewidth]{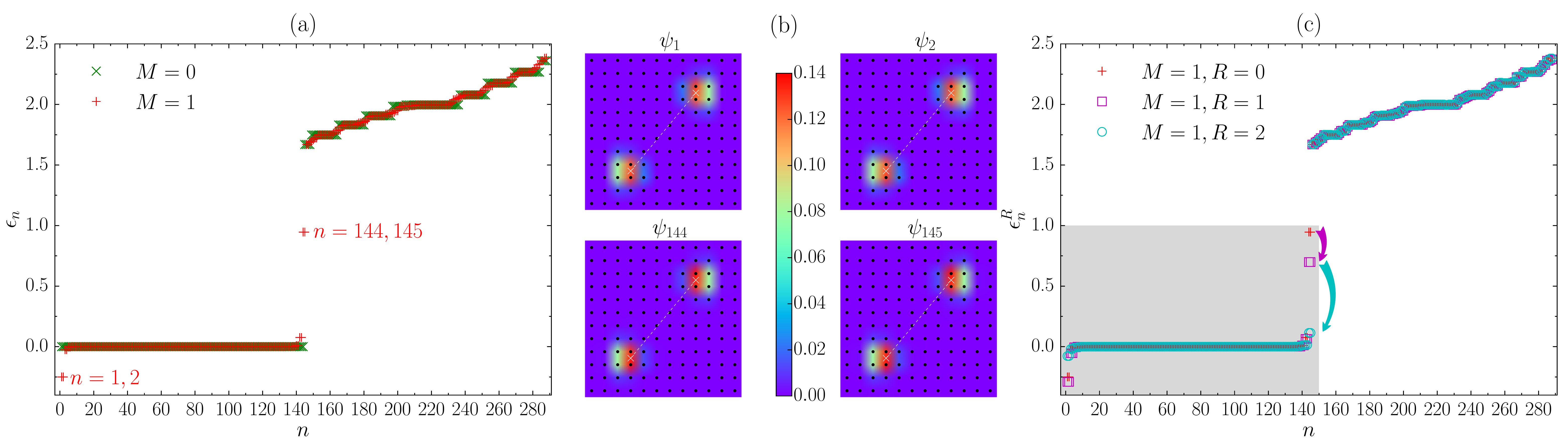}}
\caption{{\bf Single-particle spectra and defect-induced localized states for tilted branch cuts.} We study the band structure on a $L_x\times L_y=12\times12$ lattice with $\phi=1/2$. (a) The single-particle spectrum $\{\epsilon_n\}$ of $H_0$. In the absence of defects ($M=0$), $\epsilon_{1},\cdots,\epsilon_{144}$ are exactly degenerate at zero energy. With a tilted branch cut ($M=1$, white dashed line) at $(3,2.5)\rightarrow(8,8.5)$, the original band structure is distorted, with two nearly degenerate clusters $(\epsilon_1,\epsilon_2)$ and $(\epsilon_{144},\epsilon_{145})$ having the largest deviation. (b) The lattice site weight of eigenvectors $\psi_1,\psi_2,\psi_{144},\psi_{145}$ of $H_0$ for the same defects as in (a). All of them are strongly localized near the defects. The eigenstates with less energy deviation from the original band structure, for example, $\psi_3,\psi_4,\psi_{142},\psi_{143}$, are less localized (not shown here). (c) The single-particle spectrum $\{\epsilon_n^R\}$ of $H_0+V$ with $R=0$, $1$, and $2$ and the same defects as in (a). The degeneracy of $\epsilon_{1}^R,\cdots,\epsilon_{145}^R$ (shaded in gray) becomes better for larger $R$, with the flatness $0.6,1.0,8.1$ for $R=0,1,2$.}
\label{fig:SP_noty}
\end{figure*}

\begin{figure*}
\centerline{\includegraphics[width=0.85\linewidth]{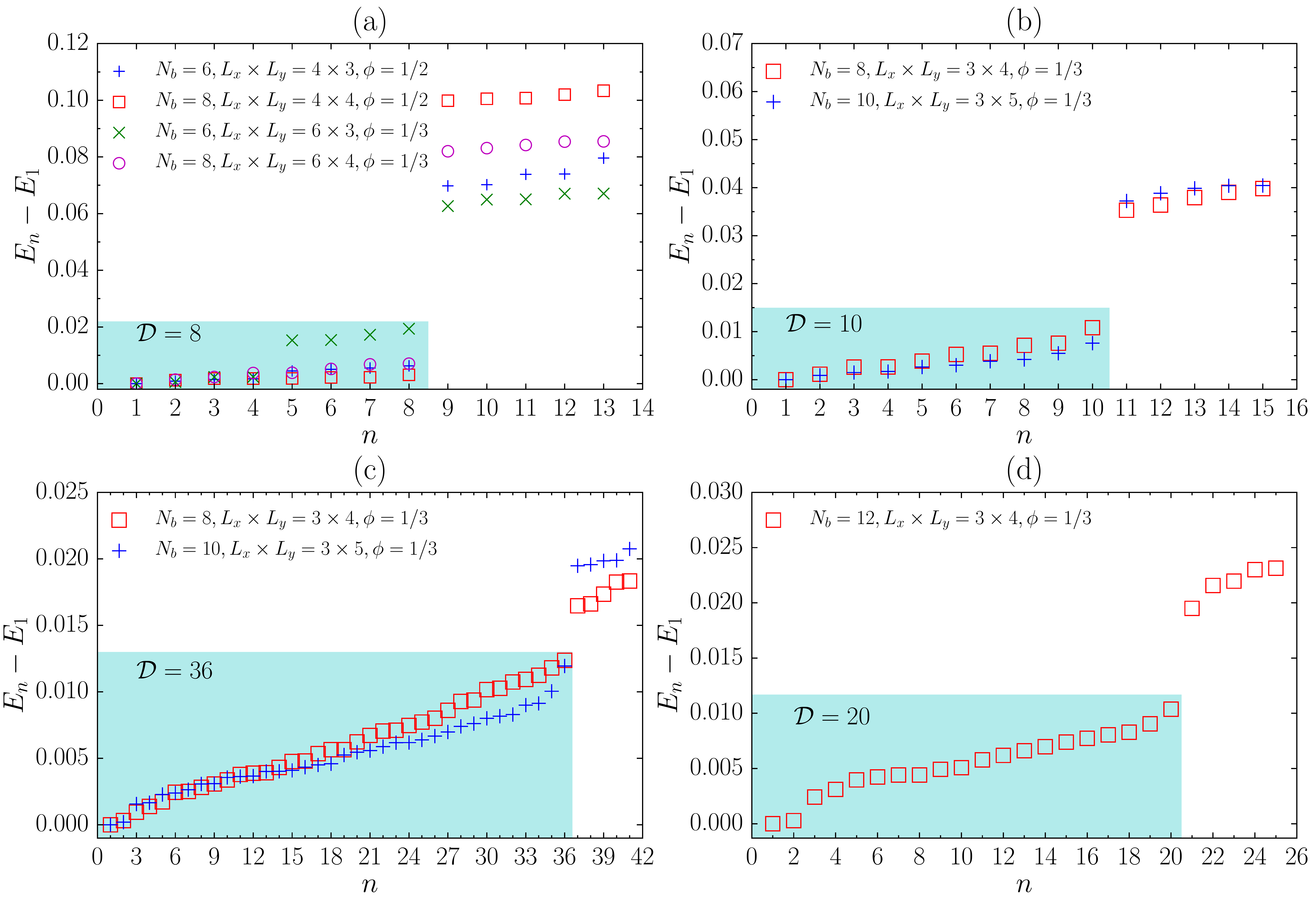}}
\caption{{\bf Many-body spectra for tilted branch cuts.} The approximately degenerate ground states, together with the degeneracy $\mathcal{D}$, are highlighted by the cyan shade. (a) $\nu=1/2$ with two branch cuts at $(0.25,0.5)\rightarrow(0.5,2),(2.25,0)\rightarrow(2.5,1.5)$ for $L_x\times L_y=4\times 3$; $(0,0.75)\rightarrow(1,2.75),(2,0.25)\rightarrow(3,2.25)$ for $L_x\times L_y=4\times 4$; $(0.25,0.5)\rightarrow(0.5,2),(3.25,0)\rightarrow(3.5,1.5)$ for $L_x\times L_y=6\times 3$; and $(0.5,0.75)\rightarrow(1.5,2.75),(3.5,0.25)\rightarrow(4.5,2.25)$ for $L_x\times L_y=6\times 4$. (b) $\nu=1$ with one branch cut at $(0.25,0.5)\rightarrow(1.75,2.5)$ for $L_x\times L_y=3\times 4$ and $(0.25,0.5)\rightarrow(1.75,3)$ for $L_x\times L_y=3\times 5$. (c) $\nu=1$ with two branch cuts at $(0.1,0.75)\rightarrow(0.4,2.75),(1.6,0.25)\rightarrow(1.9,2.25)$ for $L_x\times L_y=3\times 4$ and $(0.19,0.5)\rightarrow(0.31,3.5),(1.69,0.45)\rightarrow(1.81,3.45)$ for $L_x\times L_y=3\times 5$. (d) $\nu=3/2$ with one branch cut at $(0.5,0.5)\rightarrow(2,2.5)$ for $L_x\times L_y=3\times 4$.}
\label{fig:TD_tilted}
\end{figure*}

With tilted branch cuts, we observe a similar effect of defects on the band structure as that in the main text [Fig.~\ref{fig:SP_noty}].
Moreover, the many-body spectra of projected interactions reproduce the expected topological degeneracy for the given number of branch cuts [Fig.~\ref{fig:TD_tilted}].

\noindent{\bf Definition of particle entanglement spectra and state counting.}
PES are a useful diagnostic for topological order. For a $\mathcal{D}$-fold degenerate ground-state manifold $\{|\Psi_{\alpha}\rangle\}$ of $N$ particles, we define the PES levels $\xi$ as $\xi\equiv-\ln\lambda$, where the $\lambda$'s are the eigenvalues of the
reduced density matrix $\rho_A$ of $N_A$ particles obtained by tracing out
$N_B=N-N_A$ particles from the whole system, i.e., $\rho_A=\mathrm{Tr}_B
\rho$ with $\rho=\frac{1}{\mathcal{D}}\sum_{\alpha=1}^\mathcal{D}
|\Psi_\alpha\rangle\langle\Psi_\alpha|$. A gap in the PES is expected, below which
the number of PES levels is the same as the counting of the corresponding quasihole excitation
spectrum \cite{rbprx}, which in our case can be obtained from diagonalizing the interaction Hamiltonian of $N_b^A$ particles on the same lattice size with the same branch cuts.

\begin{thebibliography}{99}
\bibitem{tjk}
In a finite system with periodic boundary conditions, both the lattice sites and defects appear periodically. To adjust Eq.~(\ref{H0}) to periodic boundary conditions, we replace $t(z_j,z_k)$ by $\sum_{s,t=-\infty}^{+\infty}t(z_j+sL_x+\textrm{i}tL_y,z_k)e^{2\pi \textrm{i} s L_x y_j \phi}$, and in $n_{jk}$ we count all crossings between the hopping from $z_k$ to $z_j+sL_x+\textrm{i}tL_y$ and the periodically appearing branch cuts. Twisted boundary conditions can be implemented by replacing $t(z_j,z_k)$ by $\sum_{s,t=-\infty}^{+\infty}t(z_j+sL_x+\textrm{i}tL_y,z_k)e^{2\pi \textrm{i} s L_x y_j \phi}e^{-\textrm{i}s\Phi_{x}}e^{-\textrm{i}t\Phi_{y}}$.
\bibitem{Lx}
We assume $L_x$ divisible by $q$ to ensure integer number of unit cells in the $x$ direction.
\bibitem{eq3}
For a fractional quantum Hall state at filling fraction $\nu$ in the lowest Landau level on a surface of genus $g$, we have
$N_{\phi}=N/\nu-{\mathcal S}_g$, where $N_\phi$ is the number of magnetic flux quanta penetrating the surface related to the number of single-particle states $N_s$ in the lowest Landau level by $N_s=N_\phi+1-g$ \cite{wenshift}, $N$ is the number of particles, and ${\mathcal S}_g={\mathcal S}_{g=0}\times(1-g)$ is the so-called topological shift \cite{wenshift}.
For $\nu=k/2$ $\mathcal{Z}_k$ bosonic RR states, by setting ${\mathcal S}_{g=0}=2$, we have $N_s=2N/k-(1-g)$ as shown in Eq.~(\ref{Sg}).
\bibitem{projection}
Taking the two-body on site interaction as an example, the interaction after projection is
$H_{\mathrm{int}}^{\mathrm{proj}}=\sum_{m_1,m_2,m_3,m_4=1}^{2\phi L_xL_y+M}C_{m_1,m_2,m_3,m_4} a^\dagger_{m_1}a^\dagger_{m_2}a_{m_3}a_{m_4}$ with
$C_{m_1,m_2,m_3,m_4}=\sum_{i=1}^{2L_xL_y}\psi^*_{m_1,i}\psi^*_{m_2,i}\psi_{m_3,i}\psi_{m_4,i}$,
where $a^\dagger_{m}$ ($a_{m}$) creates (annihilates) a boson on the eigenstate $\psi_m$ of $H_0$, and $\psi_m=(\psi_{m,1},...,\psi_{m,2L_xL_y})$ in the lattice site basis $a_{j,\sigma}^\dagger|{\rm vaccum}\rangle$.
\end{thebibliography}
\end{document}